\documentclass[12pt]{article}
\usepackage{amssymb,amsmath,amsthm,fullpage,epsfig}
\usepackage[latin1]{inputenc}

\theoremstyle{plain}
\newtheorem{theo}{Theorem}
\newtheorem{prop}[theo]{Property}
\newtheorem{fact}[theo]{Fact}
\newtheorem{lem}[theo]{Lemma}

\theoremstyle{definition}
\newtheorem{defi}[theo]{Definition}

\theoremstyle{remark}
\newtheorem*{rem}{Remark}

\def \replace #1 by #2 ?{\fbox{#1}\marginpar{\fbox{#2}}}

\newcommand{\commentaire}[2]{
  \vskip .7cm
  \begingroup\par%
  \hbox to \hsize{\strut\llap{\bf #1\quad}\vrule\hfil\parbox{.9\hsize}{\it #2}\hfil}%
  \par\endgroup
  \vskip .7cm
}

\newcommand{\singlerule}[4]{%
      \begin{tabular}[c]{c@{\hskip .2em}c@{\hskip .2em}c}
        &#1\\
        #2&#3&#4
      \end{tabular}
}

\newcommand{\carule}[8]{%
  \begin{center}
    \small
    \begin{tabular}[c]{cccccccc}
        \singlerule#1 000
&       \singlerule#2 001
&       \singlerule#3 010
&       \singlerule#4 011
&       \singlerule#5 100
&       \singlerule#6 101
&       \singlerule#7 110
&       \singlerule#8 111
    \end{tabular}%
  \end{center}
}

\begin{document} 

\title{Cellular automata and communication complexity}

\author{Christoph D\"urr
  \thanks{Laboratoire de Recherche en Informatique, 
    Universit\'e Paris-Sud, 
    91405 Orsay, France. durr@lri.fr.
    Partially supported by the EU 5th framework programs QAIP
    IST-1999-11234 and RAND-APX IST-1999-14036, and by
    CNRS/STIC 01N80/0502 and 01N80/0607 grants.}
  \and
  Ivan Rapaport
  \thanks{Departamento de Ingenier\'{\i}a Matem\'atica
    and Centro de Modelamiento Matem{\'a}tico UMR 2071-CNRS,
    Universidad de Chile, Santiago, Chile. irapapor@dim.uchile.cl.
    Partially supported by programs FONDAP on Applied Mathematics and  
    Fondecyt 1020611.}
  \and
  Guillaume Theyssier 
  \thanks{Laboratoire de l'Informatique du Parallelisme,
    \'Ecole Normale Superieure de Lyon,
    46, All\'ee d'Italie, 69364 Lyon Cedex 07, France.  
    Guillaume.Theyssier@ens-lyon.fr}}

\maketitle

\begin{abstract} The model of cellular automata is fascinating because
very simple local rules can generate complex global behaviors. The
relationship between local and global function is subject of many
studies. We tackle this question by using results on communication
complexity theory and, as a by-product, we provide (yet another)
classification of cellular automata. \end{abstract}

\section{Introduction}


The model of cellular automata was invented in the 1950's mostly by
von Neumann as a tool to study self-reproduction
(see~\cite{neumann67}).  It was then meant both as a tool to model
real life dynamical systems and as a model of an actual computer.
Since then cellular automata are studied theoretically either as a
model of massive parallel computation or as a discrete dynamical
system. They are also studied experimentally either as a tool to model
complex natural systems ranging from economy, geology, biology,
chemistry, sociology, etc or as a framework to do simulations. For a
general introduction, see~\cite{delma}.
  
A cellular automaton is an infinite and discrete grid of cells. Each
cell contains at every time step a particular state from a finite set.
The cell state obeys a local rule, mapping its state and the state of
the neighborhood to a new cell state. This rule is applied uniformly
and synchronously to all cells of the grid. So the local rule
generates a global mapping on grid configurations, which can be quite
complex.  For example, some simple local rules give computation
universal cellular automata.

It is an important issue to understand the relationship between local
and global mappings. In this paper we view a cellular automaton as
a grid of communicating cells. During the evolution information can
flow through the whole grid. In one-dimensional cellular automata a
fixed cell divides the grid into two parts and we are interested in the
way information flows through the cell. By studying the communication
complexity of successive iterations of the local function we provide a
new way to look at the global behavior of cellular automata.

\section{Elementary cellular automata and $0-1$ matrices}

In this paper we mainly focus on elementary cellular automata (ECA)
which we define hereafter, although generalization to any
one-dimensional cellular automata (CA) is always possible and quite
straightforward.

We consider the one-dimensional cellspace, where each cell can be
either in state $0$ or in state $1$. An ECA is defined by a local
function ${f:\{0,1\}^3 \rightarrow \{0,1\}}$ which maps the state of
a cell and its two immediate neighbors to a new cell state. There
are exactly $2^{2^3}=256$ ECA and each of them is is identified with
its \emph{Wolfram number}, which is between 0 and 255 and defined as
\[
                \sum_{a,b,c\in\{0,1\}} (4a+2b+c)f(abc).
\]

Following the cellular automata's paradigm, all the cells change their
states synchronously according to the local function. This endows the
line of cells with a global dynamics whose links with the local
function are still to be understood in the general case as already
pointed out in the introduction. Let us remark however that some
simple transformations on the local function induce simple
transformations on the global dynamics: the space-symmetric ECA of $f$
is the ECA $f'$ with ${f'(a,b,c)=f(c,b,a)}$ and the state-symmetric is
the ECA $f''$ with ${f''(a,b,c)=\overline{f(\bar a,\bar b,\bar c)}}$.
Thus we consider only ECA whose Wolfram number is minimal among
its symmetries. This
leaves 88 out of 256 ECA to consider, which is more than 256/4 because
some ECA are symmetric.

To tackle the issue of local/global relationships, we study the
evolution of one cell's state after finitely many time steps. Given
that after $n$ time steps the value of a cell depends on its own
initial state and the initial states of the $n$ immediate left and $n$
immediate right neighbor cells, we define the $n$-th iteration of $f$,
${f^n: \{0,1\}^{2n+1} \rightarrow \{0,1\}}$, as  
${f^1=f}$ and for ${n\geq 2}$ as
\[
\begin{split}
f^n(x_{-n},\dots,x_{-1},x_0,x_1,\dots,x_n) = f\bigl(&
f^{n-1}(x_{-n},\dots,x_{n-2}),\\
&f^{n-1}(x_{-n+1},\dots,x_{n-1}),\\
&f^{n-1}(x_{-n+2},\dots,x_{n})\bigr).
\end{split}
\]

Notice that knowing a simple description of $f^n$ for arbitrary $n$ is
knowing the \emph{long term} (asymptotic) behavior of the whole line
of cells.

We therefore propose to measure the complexity of an ECA $f$ by the
asymptotic complexity of the functions $\bigl(f^n\bigr)$. For that
purpose, $0-1$ matrices reveal themselves to be a striking
representation. When fixing the state of the central cell among
${2n+1}$ adjacent ones to $0$ for instance, $2^{2n}$ initial
configurations are possible each leading the central cell to a
peculiar state after $n$ time steps. This can be summarized in a
square matrix $M_0^n$ of size $2^n$ defined as follows:
\[
M_0^n(i,j) := f^n(\stackrel{\leftarrow}{b_i},0,b_j)
\] where $b_k$ is
the binary representation of the integer $k$ on exactly $n$ bits, and
$\stackrel{\leftarrow}{b_k}$ its reverse representation.  
Each value $n$ defines a different matrix, and we have, for each ECA, an 
infinite family of binary matrices for $n=1,2,\ldots$.

Note that the definition of $M_0^n$ is unique up to permutation of
rows and columns. We could as well have defined $\tilde M_0^n(i,j)
:= f^n(b_i,0,\stackrel{\leftarrow}{b_j})$ for example. 

%

\begin{figure}[ht]
  \centerline{\begin{tabular}{cccc|ccccccc}
        n=&1&2&\ldots&
      n=&1&2&3&4&5&\ldots\\
      c=0&
        \epsfig{file=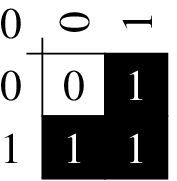,width=1cm}&
        \epsfig{file=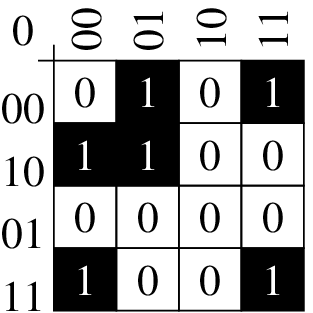,width=2cm}&\ldots&&
      \fbox{\epsfig{file=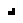,width=1mm}} &
      \fbox{\epsfig{file=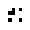,width=2mm}} &
      \fbox{\epsfig{file=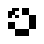,width=4mm}} &
      \fbox{\epsfig{file=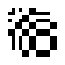,width=8mm}} &
      \fbox{\epsfig{file=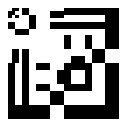,width=16mm}} &\ldots\\
      c=1&
        \epsfig{file=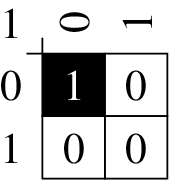,width=1cm}&
        \epsfig{file=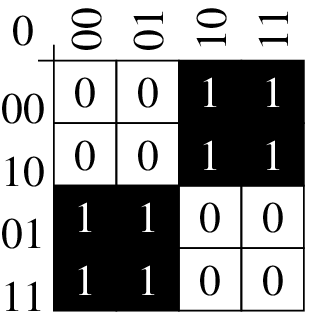,width=2cm}&\ldots&&
      \fbox{\epsfig{file=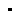,width=1mm}} &
      \fbox{\epsfig{file=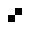,width=2mm}} &
      \fbox{\epsfig{file=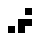,width=4mm}} &
      \fbox{\epsfig{file=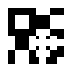,width=8mm}} &
      \fbox{\epsfig{file=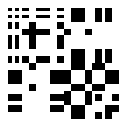,width=16mm}} &\ldots\\
    \end{tabular}}
  \caption{The two families of binary matrices $M_c^n$ of 
    Wolfram rule 54 cellular automata}
  \label{fig:054}
\end{figure}

Fixing the center cell to $0$ was arbitrary. We could as well have
chosen $1$. Therefore, any ECA $f$ defines two families of binary
matrices (see figure~\ref{fig:054}). Note that the first matrix of
each family, standing for ${n=1}$, defines completely the local function.
One can think of these matrices as seeds for the families.

\begin{figure}[ht]
  \centerline{\epsfig{file=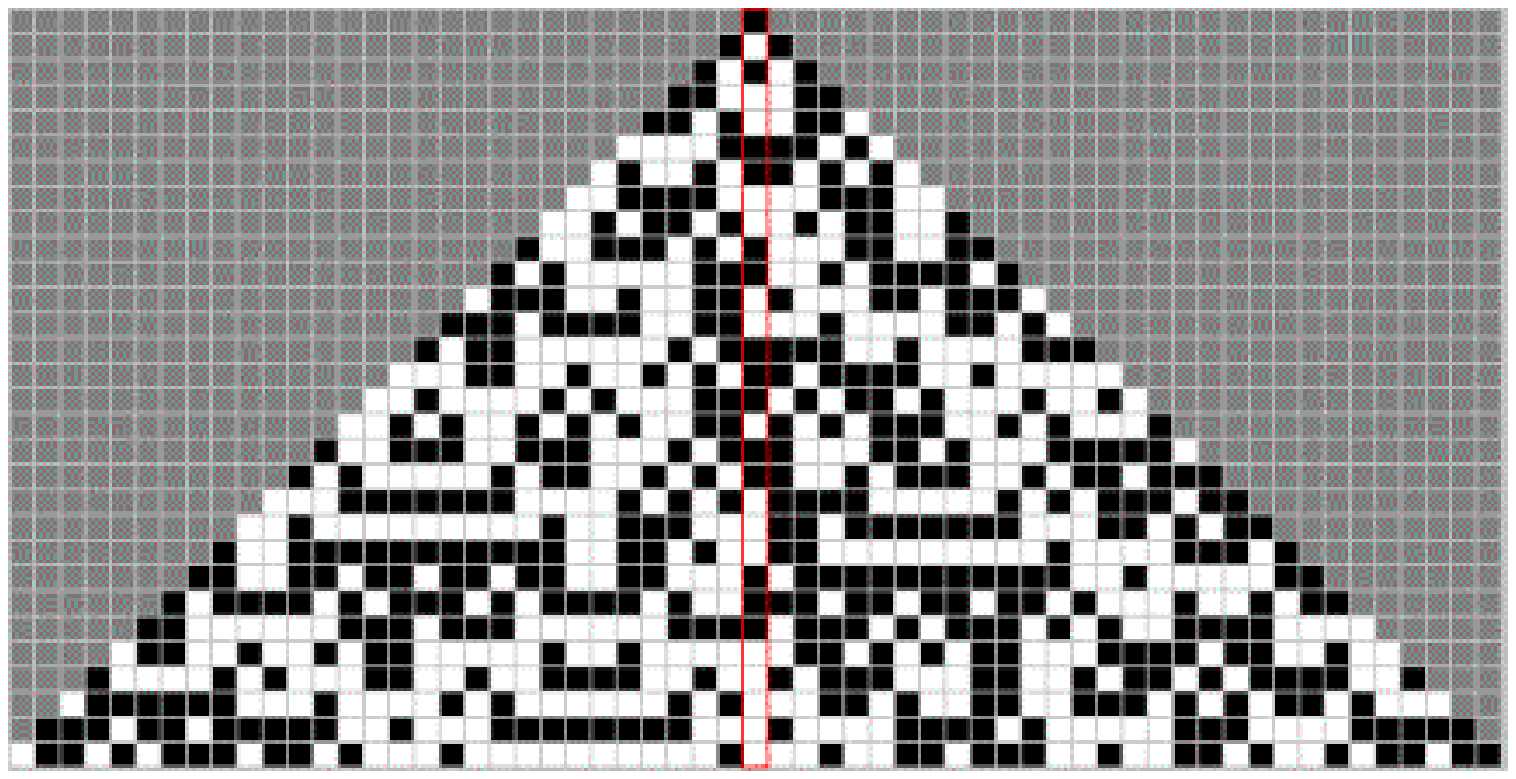,width=6cm}
    \hspace{1cm}
    \fbox{\epsfig{file=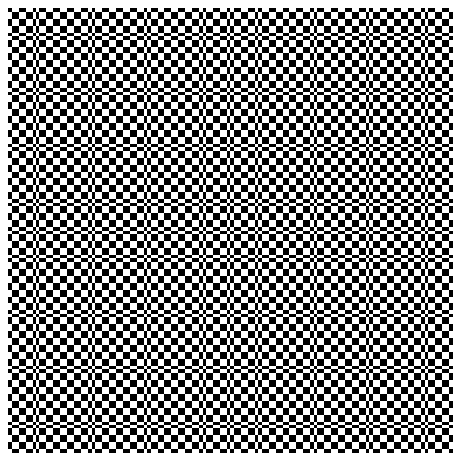,width=3cm}}
  }
  \caption{A space time diagram for rule 105 (left) and
    one matrix of its families (right).
    In the diagram every row is a configuration and time goes upward.
    It shows only those cells, on which the upper center cell depends.}
  \label{fig:105}
\end{figure}

The matrices can in some cases ease the understanding of the global
behavior. In figure~\ref{fig:105} we show the space time diagram of
rule 105 for some arbitrary configuration, and on the right the matrix
$M_0^5$. In contrast with the space-time diagram, the matrix looks
simple, and indeed there is a small description of the \emph{additive}
rule 105 (which is given later in the paper). We should emphasize
that the space-time diagram shows the evolution of only a single
configuration, while the matrix covers all configurations.

Different measures on the matrices are possible in order to analyze the 
underlying ECA.
Among them we choose the simplest one: the number of different rows.
To be precise we define $d_n$ as the
maximum of the $4$ following integers: number of different rows of
$M_0^n$, number of different columns of $M_0^n$, number of different
rows of $M_1^n$, number of different columns of $M_1^n$.

\section{Experimental measuring}

Since the family of matrices of a given ECA defines the global and long
term behavior, we can express $d_n$ as a known function of $n$ only once
we understood the global behavior. However in some cases, seeing the
matrices helped us to understand the global function.

\begin{figure}[ht] 
  \centerline{\epsfig{file=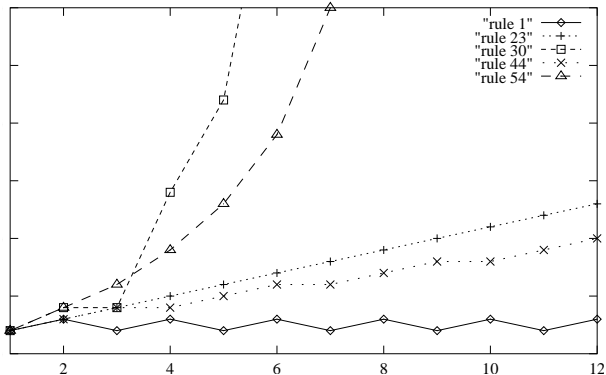,width=8cm}}
  \caption{Different sequences $d_n$.}
  \label{fig:dn}
\end{figure}

We did brute force computations in order to compute $d_n$ for
$n\in\{1,\ldots,12\}$ and for all ECAs.
The complete results are shown in the web page www.lri.fr/\verb|~|durr/CACC/.
Figure~\ref{fig:dn} plots $d_n$ for different rules.
We obtain quite different sequences, which we classify as follows:
\begin{description}
\item[Bounded:]
  there is ${b\in\mathbb N}$ such that ${\forall n\ :\ d_n\leq b}$.


\item[Linear:] there are values ${a_0\in\mathbb N}$ and
  ${a_1\in\mathbb Q}$ such that $d_n=\lfloor a_1n\rfloor +a_0$,
  for all $n\geq n_0$ for some fixed value $n_0$.
\item[Other:] in this class we put rules where non of above applies.
  In some cases $d_n$ seems to be bounded by a polynomial in $n$, 
  and in some cases $d_n$ seems to be exponential.
\end{description}

We want to emphasize that this classification is mainly experimental.
Most of the time we don't have mathematical evidence for determining
whether a rule belongs to one class rather than to another.
Table~\ref{table:dn} shows the classification of all rules (again, only
up to symmetric transformations which preserve $d_n$).

The following sections are devoted to the few ECA where we were able
to give a closed formula for $d_n$ making their classification
rigorous.

\begin{table}[ht]
  \begin{description}
  \item[Bounded:] 0, 1, 2, 3, 4, 5, 7, 8, 10, 12, 13, 15, 19, 24, 27,
    28, 29, 32, 34, 36, 38, 42, 46, 51, 60, 71, 72, 76, 78, 90, 105,
    108, 128, 130, 136, 138, 140, 150, 154, 156, 160, 170, 172, 200,
    204
  \item[Linear:] 11, 14, 23, 33, 35, 43, 44, 50, 56, 58, 77, 132, 142,
    152, 168, 178, 184, 232
  \item[Other:] 6, 9, 18, 22, 25, 26, 30, 37, 40, 41, 45, 54, 57, 62,
    73, 74, 94, 104, 106, 110, 122, 126, 134, 146, 164
  \end{description}
  \caption{A classification of ECAs}
  \label{table:dn}
\end{table}

\section{Communication complexity}

The communication complexity framework appears as an extremely useful
tool for calculating $d_n$. The communication complexity theory studies
the information exchange required by different actors to accomplish a
common computation when the data is initially distributed among them. To
tackle that kind of questions, \textsc{A.C.~Yao} \cite{yao79} suggested
the two-party model: two persons, say Alice and Bob, are asked to
compute together the values taken by a function $f$ of $2$ variables
($x$ and $y$ taking values in a finite set), Alice always knowing the
value of $x$ only and Bob that of $y$ only. Moreover, they are asked to
proceed in such a way that the cost --- the total number of exchanged
bits --- is minimal in the worst case. Now different restrictions on
the communication protocol lead to different communication complexity
measures.

\begin{defi}[Many round communication complexity]
  The many round communication complexity $CC(f)$ of a function $f$ is
  the cost of the best protocol for $f$.
\end{defi}

\begin{defi}[One-way communication complexity] 
  A protocol is AB-one-way
if only Alice is allowed to send information to Bob, and Bob has to
compute the function solely on its input, and the received information.
The AB-one-way communication complexity is the worst case number of
bits Alice needs to send. BA-one-way complexity is defined in the same
manner.
  Finally, the one-way complexity $C(f)$ of $f$ is the maximum of its
  AB-one-way and BA-one-way complexities.
\end{defi}

Whereas most studies concern the many round communication complexity, we
focus only on the one-way communication complexity. In terms of cellular
automata it will permit us to measure the amount of information which
have to flow from one side to another. Also from a practical point of
view, the former measure is extremely difficult to compute for most
functions, while the last measure is quite easy as shown by the next
fact.



\begin{fact}[\cite{kushilevitz97}]
  \label{fact:oneway}
  Let $f$ be a binary function of ${2n}$ variables and
  $M_f\in\{0,1\}^{2^n\times2^n}$ its matrix representation, defined by
  $M_f(x,y)=f(xy)$ for $x,y\in\{0,1\}^n$. Let $d(M_f)$ be the
  maximum of the number of different rows and the number of different
  columns in $M_f$.
  We have
  \[C(f) = \left\lceil\log_2\bigl(d(M_f)\bigr)\right\rceil.\]
\end{fact}
\begin{proof}
  Let be a AB-one-way protocol, where Alice knows $x\in\{0,1\}^n$ and
  Bob $y\in\{0,1\}^n$.  Suppose Alice sends to Bob at most $k$ bits
  which depend solely on $x$, say by some mapping
  ${g:\{0,1\}^n\rightarrow \{0,1\}^k}$. Then since Bob knows $f(x,y)$
  only from $y$ and $g(x)$, we must have $f(x,y)=f(x',y)$ for all
  $x,x'$ with $g(x)=g(x')$.  In terms of matrix representation it
  means that the rows in $M_f$ indexed by $x$ and $x'$ are the same.
  So we must have ${k\geq
    \left\lceil\log_2\bigl(d(M_f)\bigr)\right\rceil}$.
  
  Conversely, ${\left\lceil\log_2\bigl(d(M_f)\bigr)\right\rceil}$ are
  sufficient for Alice: knowing $M_f$ she only has to say to Bob the
  group of identical rows the current entry belongs to.
\end{proof}

Then, up to a $\log$ transformation, the complexity measure $d_n$
appears to be precisely the exchanged information amount required for
Alice and Bob to compute $f^n$ when Alice has the $n$ left cells and Bob
has the $n$ right cells, both knowing the value of the central cell and
maximizing over the scenario where only Alice is allowed to talk, and
the scenario where only Bob is allowed to talk.


%

\section{ECA with bounded complexity}

We will now give formal proofs for some ECA to be in the bounded
complexity class.

\begin{defi}
  An ECA is \emph{nilpotent} if it converges to a unique configuration
  from any initial configuration in finite time.
\end{defi}

Given that an ECA $f$ is nilpotent if and only if there is $n_0$ such
that $f^n$ is constant for all $n\geq n_0$, it is clear that a
nilpotent ECA will have $d_n=1$ for large enough $n$: no communication
is needed between Alice and Bob to compute the final state of the
central cell.

There is a natural condition generalizing nilpotency, which
can be used to prove bounded complexity for ECA.

\begin{defi}
  An ECA $f$ has a \emph{limited sensibility} if the number of cells
  $f^n$ actually depends on is bounded by a constant independent of
  $n$.  Formally there is a constant $c$ such that $\forall n$ $\exists
  w \in \{0,1\}^{2n+1}$ with $\sum_{i=1}^{2n+1} w_i\leq c$ and
  $f^n(u)=f^n(u\wedge w)$ for all $u\in\{0,1\}^{2n+1}$ where $\wedge$ is
        the bitwise boolean and. 

  The ECA has \emph{half-limited sensibility} if the condition above
  holds for the weaker condition $\sum_{i=1}^n w_i\leq c$ or
  equivalently $\sum_{i=n+1}^{2n+1} w_i \leq c$. 
\end{defi}

Clearly if $f$ has limited sensibility $c$ at most $c$ bits need to be
exchanged between Alice and Bob. We show now that the half-limited
sensibility is enough to achieve bounded one-way communication
complexity.

\begin{lem}
  An ECA with half-limited sensibility is in the bounded complexity
  class.
\end{lem}
\begin{proof}
  Without loss of generality, let us assume that Alice knows the left
  cells and that the sensibility of $f^n$ is limited by $c$ on the first
  $n$ cells. There is a trivial AB-one-way protocol of cost $c$. We
  give a BA-one-way protocol of cost $2^c$, which is worse but still
  constant. Bob successively guesses each possible value of Alice's
  sensible cells and sends the list of the corresponding values for
  $f^n$. Then Alice select among this list, the entry corresponding to
  the actual values of its sensible cells.
\end{proof}

An interesting example is rule~60. It has sensibility $0$ on the second
half, and unlimited sensibility on the first half, as it computes the
parity of the first $n+1$ cells, whenever $n+1$ is a power of $2$.

Unfortunately, it is undecidable to determine whether a given CA has a
limited sensibility (by reduction from the nilpotency problem which is
undecidable~\cite{kari92}). Except some particular examples, we can
only guess if a ECA has limited sensibility based on brute force
computation for small values of $n$.

However there is a decidable property, which is sufficient for a ECA
to be in the bounded class.

\begin{defi}
  \label{def:add}
  An ECA $f$ is \emph{additive} is there are 
  binary operators $\oplus$ and
  $\otimes$ (not necessarily distinct), and a  neutral element $e$
  such that for all $x$, ${x\oplus e  = e \oplus x = x}$ and with
  \begin{equation*}
    \forall x,y,z,x',y',z'\quad\left\lbrace
    \begin{aligned}
      f(x\oplus x',y\oplus y',z\oplus z') &= f(x,y,z)\otimes f(x',y',z')\\
      f(x\otimes x',y\otimes y',z\otimes z') &= f(x,y,z)\oplus
      f(x',y',z')
    \end{aligned}\right.
  \end{equation*}
\end{defi}

\begin{rem}
  The previous definition can naturally be extended to longer operator
  chains.  However, as far as ECA are concerned, $2$ operators are
  sufficient to capture all additive rules.
\end{rem}

Additivity is preserved by iterations of the local rule, as stated in the
following lemma which can be proved straightforwardly by induction on $n$.
\begin{lem}
  \label{lem:add}
  For all integer $n$ we have $\forall u,u'\in\{0,1\}^{2n+1}$:
  \begin{align*}
    f^n(u\oplus u') &= &
    \begin{cases}
      f^n(u)\otimes f^n(u')& \text{if $n$ is odd,}\\
      f^n(u)\oplus f^n(u')& \text{otherwise,}
    \end{cases}\\
    f^n(u\otimes u') &= &
    \begin{cases}
      f^n(u)\otimes f^n(u')& \text{if $n$ is even,}\\
      f^n(u)\oplus f^n(u')& \text{otherwise.}
    \end{cases}
  \end{align*}
  where $\oplus$
  and $\otimes$ act bitwise on bitstrings.
\end{lem}

\begin{prop}
  If $f$ is additive, $d_n$ is bounded by 2.
\end{prop}
\begin{proof}
  Let $f$ be additive for $\oplus$ and $\otimes$ with  the neutral element $e$.
  We give a one-way protocol computing $f^n$, where Alice sends a single bit
  to Bob, who can then compute the function.
  Bob could have started as well in this protocol.
  Let $c$ be
  the state of the central cell and $u$ and $v$ the states of
  the
  $n$ left cells and $n$ right cells. Alice knows $u,c$; Bob knows
  $v,c$; and the goal is to compute $f^n(u,c,v)$.  The protocol is the
  following:
  \begin{enumerate}
  \item Alice computes 
        the single bit ${b=f^n(u,c,\underbrace{e\cdots e}_n)}$ and sends
    it to Bob;

  \item then Bob computes 
    \[
    \begin{cases}
      b\oplus f^n(\underbrace{e\cdots e}_n,c,v) &\text{if $n$ is
        even}\\
      b\otimes f^n(\underbrace{e\cdots e}_n,c,v) &\text{if $n$ is odd.}
    \end{cases}
    \]
  \end{enumerate}
  By Lemma~\ref{lem:add} the protocol is correct.
\end{proof}

As an example we consider the additive rule 105. In the following
table we write its local function $f(x,y,z)$ above all combinations of
$x,y,z$. By the way, note that by definition of Wolfram numbers, the
first row contains the number 105 in reverse binary notation.

{\carule 10010110}

The local function can be written as
$$f(x,c,y)=x\oplus c \oplus y \oplus 1,$$
where $\oplus$ is the
exclusive or. Given that $\oplus$ is associative, commutative and
admits a neutral element ($0$), the rule clearly fits
definition~\ref{def:add}. This explains why the matrices $M_c^n$ have
only two different rows or columns, as depicted in figure~\ref{fig:105}.

These observations permit us to refine the class of ECA having bounded
complexity, and distinguish the following cases. Note that the first
subclass is rigorous as it is based on proven properties of the local
function, while limited sensibility is mostly based on brute force
computation. The last subclass contains ECA for which we were not
able to give a general reason for their membership to this class,
although some of them are easy to understand individually (for
instance, rule $32$).

\begin{description}
\item[bounded by additivity]
  15, 51, 60, 90, 105, 108, 128, 136, 150, 160, 170, 204
\item[bounded by limited sensibility] 0, 1, 2, 3, 4, 5, 8, 10, 12,
  19, 24, 29, 34, 36, 38, 42, 46, 72, 76, 78, 108, 138, 200
\item[bounded by half-limited sensibility] 7, 13, 28, 140, 172
\item[bounded for any other reason] 27, 32, 130, 156, 162
\end{description}

\section{ECA with linear complexity}

\begin{figure}[ht]
  \begin{tabular}{llll}
    \fbox{\epsfig{file=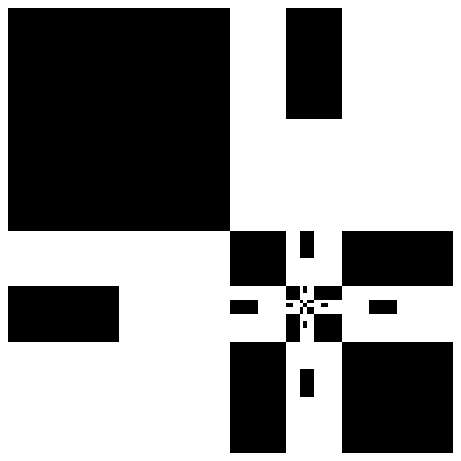,width=3cm}} &
    \fbox{\epsfig{file=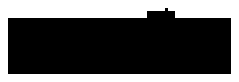,width=3cm}} &
    \fbox{\epsfig{file=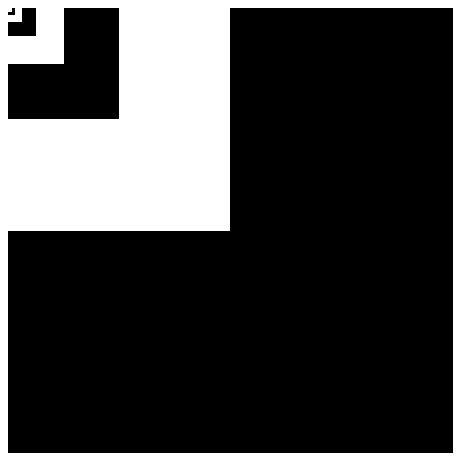,width=3cm}} &
    \fbox{\epsfig{file=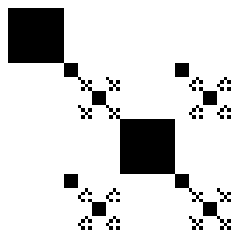,width=3cm}} \\
    rule 33 & rule 44 & rule 50 & rule 164\\
    \fbox{\epsfig{file=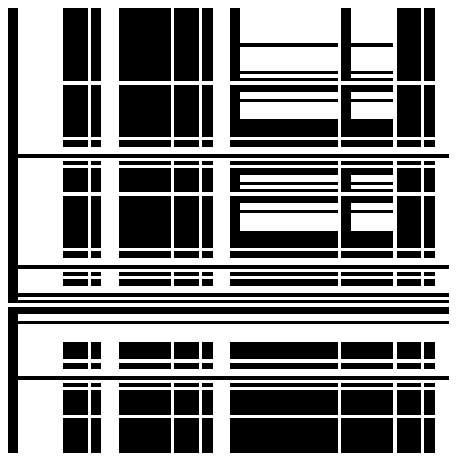,width=3cm}} &
    \fbox{\epsfig{file=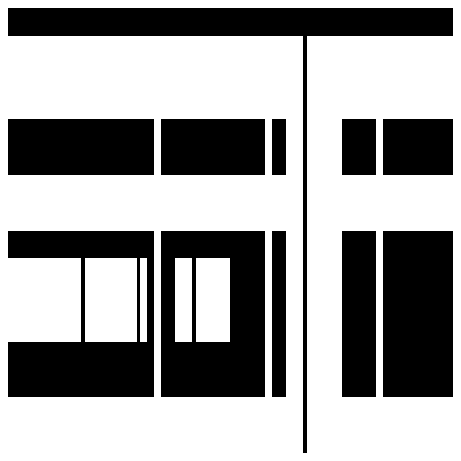,width=3cm}} &
    \fbox{\epsfig{file=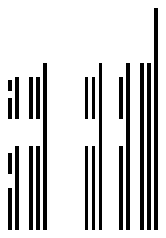,width=3cm}} &
    \fbox{\epsfig{file=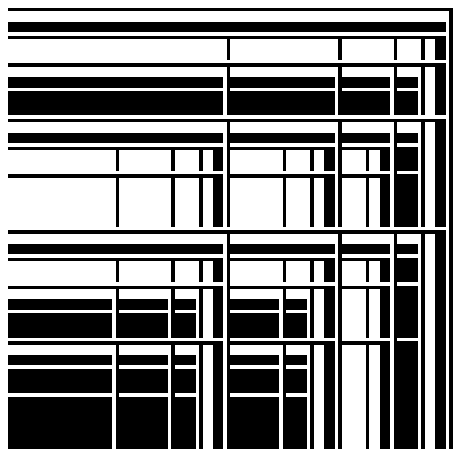,width=3cm}}\\
    rule 14 & rule 35 & rule 168 & rule 184
  \end{tabular}
  \caption{Matrices of some ECA with linear complexity.  
    The first row shows matrices with striking fractal structure.}
  \label{fig:linear}
\end{figure}

The case of linear complexity illustrates very nicely the relationship
between communication complexity and ECA. We would like to emphasize
on the fractal structure of the matrices for some rules, see
figure~\ref{fig:linear}. For those matrices the number of different rows
is logarithmic in the size of the matrix, which makes it linear in $n$.

As an example we consider rule 132:

{\carule 00100001}

\begin{figure}[ht]
  \centerline{\epsfig{file=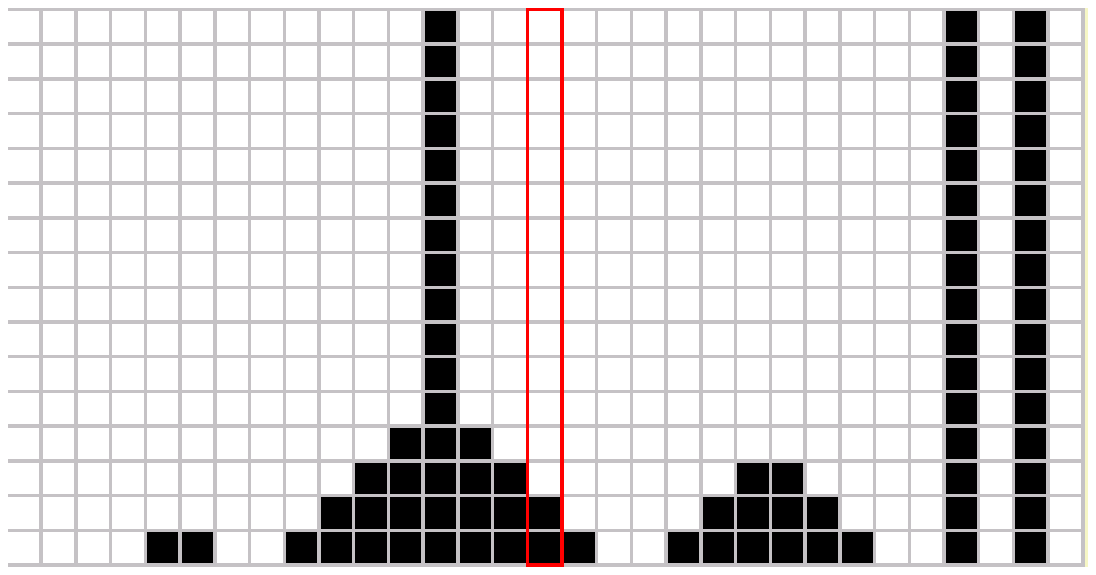,width=6cm}
    \hspace{1cm}
    \fbox{\epsfig{file=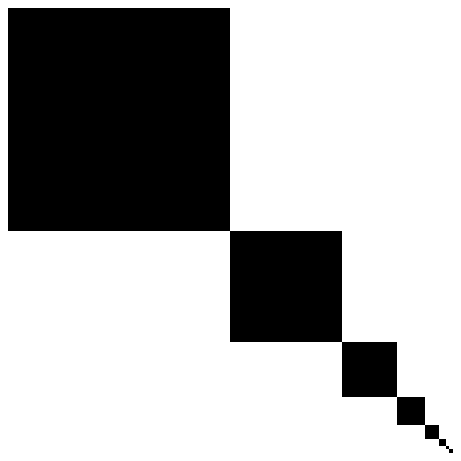,width=3cm}}
  }
  \caption{A space time diagram for rule 132 (left) and
    one matrix of its families (right).  Note that in this matrix 
    the lower right quadrant
    has the same structure as the matrix itself.}
  \label{fig:132}
\end{figure}

Figure~\ref{fig:132} shows the fractal structure of its matrices. The
space time diagram gives an explanation. Any block consisting of several
1's, shrinks at every time step by 1 at each end, and either vanishes or
remains a single 1, depending on the parity of its length.

\begin{lem}
  For rule 132 we have ${d_n=n+1}$.
\end{lem}

\begin{proof}
  For an upper bound we give a very simple communication protocol. If
the center is $0$, no communication at all is needed, as the answer will
always be $0$. Now suppose the center is $1$. It is part of a block
consisting of $k$ cells on Alice's side, and of $l$ cells on Bob's side,
where $0\leq k,l \leq n$. Now Alice sends $k$ to Bob, and Bob answers
$1$ if $k=l$ and $0$ otherwise. The protocol is correct, since after
$n$ steps the center will remain $1$ only if it is in the center of an
even length block or if it is distant by at least $n$ to each end of
the block. For the protocol, Alice needs to send a number out of $n+1$
different values. It would be the same if Bob starts first. Therefore
$d_n \leq n+1$.

For the lower bound, we give a submatrix in $M_0^n$ which is the
identity. This submatrix is of dimension $(n+1)\times(n+1)$, which
will show that there are at least $n+1$ different rows and at least as
must different columns. Let $R_k$ be the row in $M_0^n$ corresponding
to $0^{n-k}1^k$, and let $C_l$ be the column corresponding to
$1^l0^{n-l}$.  The submatrix made up of the intersection of rows $R_k$
and columns $C_l$ for $0\leq k,l \leq n$ is the identity matrix.
\end{proof}

This example is also interesting because the number of different rows in
$M_c^n$ is $1$ if $c=0$ and $n+1$ if $c=1$, while for most rules 
the number of different rows seems to be the same (or at least similar)
for the two families of matrices.


\begin{figure}[ht]
  \centerline{\epsfig{file=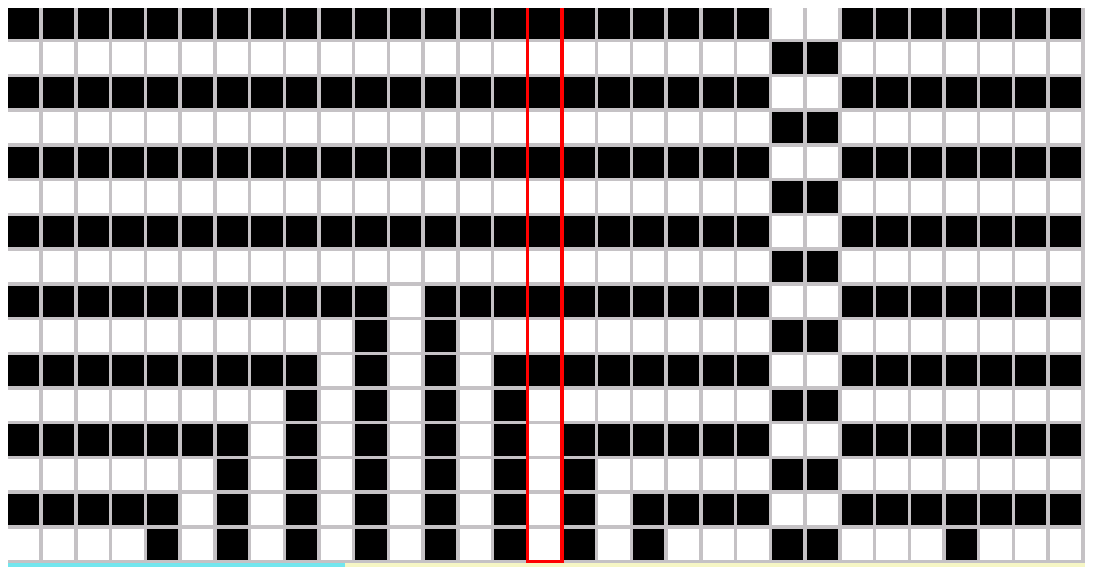,width=6cm}
    \hspace{1cm}
    \fbox{\epsfig{file=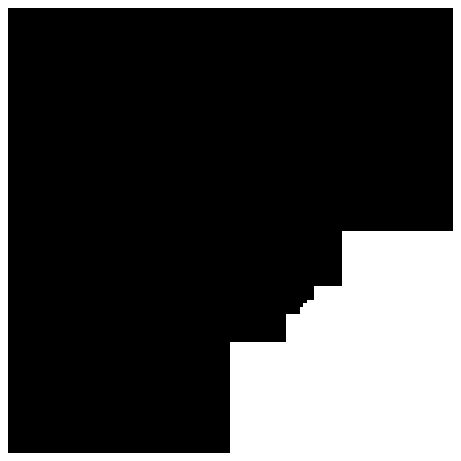,width=3cm}}
  }
  \caption{A space time diagram for rule 23 (left) and 
    one matrix of its families (right).  Note that there
    is a square in the matrix, smaller by 1/4 in height and width
        which has the same structure as the matrix itself.}
  \label{fig:023}
\end{figure}

Another example is rule 23:
{\carule 11101000}

In this rule, cells become alternatively 0 and 1 with exception of those
being inside a block of alternating 0's and 1's. Since the neighboring
cells of such a block alternate their states, a block shrinks at both
ends by 2 cells every second time steps. One can apply the same method
to prove $d_n=n+1$.

\section{ECA with other complexity}

The ECA of this class are not well understood yet. For example, it
contains rules 110 and 54, which are conjectured to be computation
universal.\footnote{Rule 110 has been proven universal in some sense
  by \textsc{M.~Cook} who presented his results during the CA98
  workshop at Santa Fe Institute. The result is also presented in the
  book ``A New Kind of Science'' by \textsc{S.~Wolfram} . But as far
  as we know a complete and detailed proof doesn't appear in any
  reference.} 
However for larger cell states we were able to prove polynomial and
even exponential complexity for some CA.

The following CA with $3$ states has quadratic complexity and the
construction can be generalized for more states to achieve complexity
$\Theta(n^k)$ for any $k$.

Let be the state set ${\{0,1,2\}}$. The cellular automata is
defined for all $i,j,k$ by
\[
	f(i,j,k) = \left\{ \begin{array}{ll}
		j & \text{if }j=0 \text{ and } i=k,\\
		\max\{i,j,k\} &\text{otherwise}.
	\end{array}\right.
\]

\begin{figure}[ht]
  \centerline{\epsfig{file=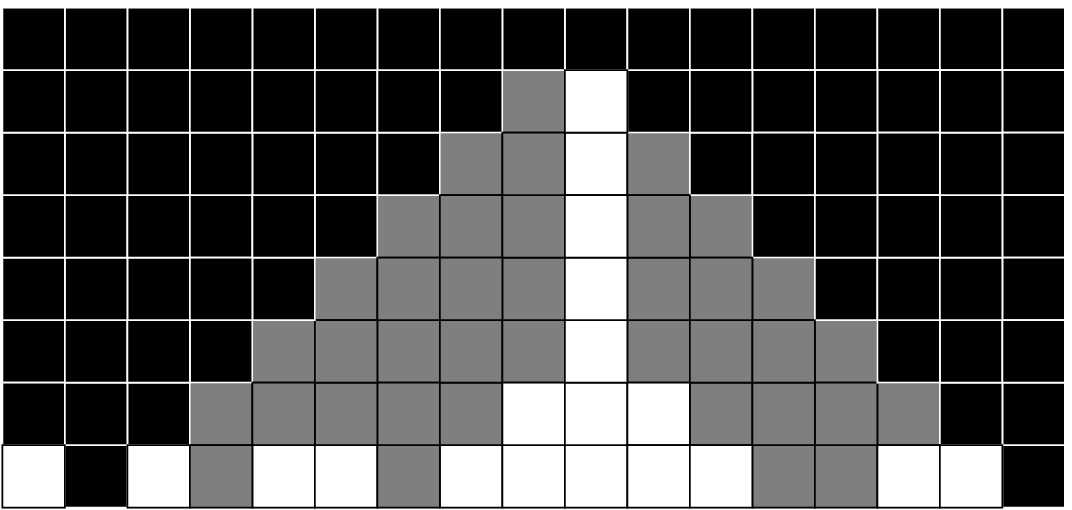,width=6cm}
    \hspace{1cm}
    \fbox{\epsfig{file=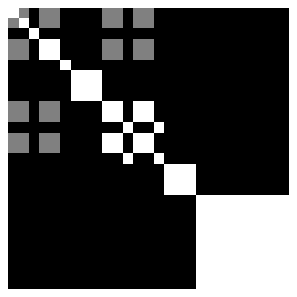,width=3cm}}
  }
  \caption{A space time diagram for the 3-state CA, where
    0=white, 1=gray, 2=black
    (left) and one matrix $M_0^n$ of its families (right).}
  \label{fig:n_2}
\end{figure}

Note that each state is quiescent. The global behavior can be explained
like this (see figure~\ref{fig:n_2}). Every cell different from $0$
tends to expand in both directions. Whenever a $1$-expansion and a
$2$-expansion meet, the $2$-expansion overrules the former. There is
a single situation where a $0$ cell can remain $0$, it is when two
expansions of the same state reach at the same time the left and right
neighbor.

To make this statement formal, consider the cell at position $0$. Let
$r_2$ be the position of the first state-2-cell on the right between
$1$ and $n$, and $\infty$ if there is no such cell. Let $r_1$ be the
position of the first state-1-cell between $1$ and $r_2-1$ and $\infty$
if there is no such cell. The positions $l_1,l_2$ are defined in the
same manner for the left neighborhood. Now we can completely determine
the cell state after $n$ steps.
\begin{itemize}
\item
	If the cell was in state $2$, then it remains $2$ forever.
\item
	If the cell was in state $1$, then it remains $1$ after $n$ steps
	if and only if $l_2<-n$ and $n<r_2$.
\item
	If the cell was in state $0$, then it remains $0$ after $n$ steps
	if and only if ($l_2<-n$ and $n<r_2$ or $-l_2=r_2$) and
		       ($l_1<-n$ and $n<r_1$ or $-l_1=r_1$).
\end{itemize}

From these observations we can determine its complexity.

\begin{lem}
  The complexity of the CA defined above is $d_n\in\Theta(n^2)$.
\end{lem}
\begin{proof}
  We will construct $\Omega(n^2)$ different rows in the matrix
  $M_0^n$, which will suffice for the lower bound by symmetry of
  the matrix. For every $0\leq j\leq i\leq n$ let be $R_{i,j}$ the
  row corresponding to the left neighborhood \[\underbrace{2\cdots
  2}_{n-i}\underbrace{1\cdots 1}_{i-j}\underbrace{0\cdots 0}_j.\] Let
  be $C_{i,j}$ the column corresponding to the similar (reversed) right
  neighborhood. Then row $R_{i,j}$ is $0$ only at column $C_{i,j}$,
  showing that all rows $R_{i,j}$ are different.

  Conversely for the upper bound, it is clear from the above description
of the rule that a correct one-way communication protocol can be
achieved with complexity ${2\lceil\log_2 (n+1)\rceil}$, since Alice
needs only to send the numbers $l_1,l_2$ to Bob, while each can have
only $n+1$ different values.
\end{proof}

We show now a very well known and simple CA, which has exponential
complexity.

Let be state set $\{0,1,\tilde0,\tilde1\}$. We define the cellular
automata $f$ such that $f^n(u,1,v)$ compares the strings $u$ and
$v$, with $u\in\{0,1\}$ and $v\in\{\tilde0,\tilde1\}$. Let be
$x,c,y\in\{0,1\}$ and $z\in\{0,1,\tilde0,\tilde1\}$, then
\begin{eqnarray*}
  f(z,\tilde c, \tilde y) = \tilde y && \text{(shift left to the center)}\\
  f(x, c, y) = x && \text{(shift right to the center)}\\
  f( x, 1, \tilde y) = \left\{\begin{array}{ll}
      1 & \text{if }x=y\\
      0 & \text{if }x\neq y
    \end{array}\right. && \text{(a difference makes the center 0)}
\end{eqnarray*}
For all other values $f$ is $0$. 

\begin{lem}
  The complexity $d_n$ of the CA defined above is exponential in $n$.
\end{lem}
\begin{proof}
Clearly $M_1^n$ contains at least
$2^n$ different rows (among $4^n$): for every row indexed by ${u\in\{0,1\}^n}$
there is a single entry $1$, at a column $v$, where $v$ is the reverse
dual of $u$, and any other row contains only $0$.
\end{proof}


\section{Future directions}

The purpose of this paper was to show a relationship between 
cellular automata and communication complexity.  For this we 
deliberately made simple choices: 
elementary cellular automata and particular one-way communication.
 Our approach can be generalized in several ways:

\begin{itemize}
\item We partitioned evenly the $2n+1$ cells between Alice and Bob, and also fixed the center cell.  This does not take into account asymmetric behavior for some CA.
   When $f^n$ is fixed,
  we can define a matrix representation $M_p^n$ of dimension
  $2^p\times 2^{2n+1-p}$ (for ${1\leq p\leq 2n+1}$) such that
  $M_p^n(u,v)=f(u\stackrel{\leftarrow}{v})$, \emph{i.e.} where Alice knows
  the $p$ leftmost cells and Bob the $n-p$ rightmost. Then an interesting
  complexity measure is:
  \[R_n = \max_p |rows(M_p^n)|,\]
  where $|rows(\cdot)|$ stands for the number of different rows.
  The value of $p$ where the maximum is reached is non trivial and,
  while clearly connected to the global behavior, hard to predict from
  the local rule. A striking example is ECA rule 7 which has a bounded
  $d_n$ complexity due to its half-limited sensibility but has a
  linear $R_n$ complexity, maximum being reached around
  ${p=\frac{n}{3}}$, as indicated by brute force computations. 

  Following this idea, we could as well define $R_n$ symmetrically on
  rows and columns as follows:
  \[R_n = \max_p \max\left\{|rows(M_p^n)|,|cols(M_p^n)|\right\}.\]

\item The link between communication complexity and matrix
  representations is not reduced to Fact~\ref{fact:oneway}. Actually,
  many round communication complexity is lower-bounded by the log of
  the rank of the matrix representation (see~\cite{kushilevitz97}).
  Moreover, it is conjectured that a poly-log of the rank is also an
  upper-bound.

\item Finally, considering the more general framework of all
  one-dimensional CA (with any radius and any state set), it is
  interesting to note that a classification based on the asymptotic
  behavior of $d_n$ has some very reassuring properties such as:
  \begin{itemize}
  \item the class of a CA $A$ is ``higher'' than the class of any of
    its sub-CA ;
  \item the complexity of the Cartesian product $A\times B$ is the
    product of the complexities of $A$ and $B$.
  \end{itemize}
\end{itemize}
Another important issue of course, is to give mathematical rigorous proofs
for the complexity of some ECA, which however is at least as difficult
as understanding completely the global behavior.



\section{Acknowledgment}

We would like to thank Nicolas Ollinger for helpful discussions.

\bibliographystyle{alpha} 
\bibliography{cacc}

\end{document}